\documentclass{amsart}
\usepackage{amssymb}
\usepackage{latexsym}

\newcommand{\RR}{{\mathbb R}}

\newcommand{\NN}{{\mathbb N}}

\newtheorem{theorem}{Theorem}

\newtheorem{lemma}{Lemma}[section]
\newtheorem{prop}[lemma]{Proposition}
\newtheorem{coro}[lemma]{Corollary}

\renewcommand\qedsymbol{$\Box$}

\newcommand{\tr}{{\mathrm{tr}}}
\newcommand{\osc}{{\mathrm{osc}}}

\begin{document}
\sloppy
\title[A generalization of Gordon's theorem]{A generalization of Gordon's theorem and applications to quasiperiodic Schr\"odinger operators}

\author[D.~Damanik, G.~Stolz]{David Damanik$\,^{1}$, G\"unter Stolz$\,^{2}$}
\noindent\thanks{$^1$ Department of Mathematics 253--37, California Institute of Technology, Pasadena, CA 91125, U.S.A.\ and Fachbereich Mathematik, Johann Wolfgang Goethe-Universit\"at, 60054 Frankfurt, Germany. Research supported by the German Academic Exchange Service through Hochschulsonderprogramm III (Postdoktoranden). E-mail: \mbox{damanik@its.caltech.edu}\\[1mm]
{\indent}$^2$ Department of Mathematics, University of Alabama at Birmingham, Birmingham, AL 35294, U.S.A. Research partially supported by NSF Grant DMS 9706076. E-mail: \mbox{stolz@math.uab.edu}\\[1mm]
}

\maketitle

\begin{abstract}
We prove a criterion for absence of eigenvalues for one-dimensional Schr\"odinger operators. This criterion can be regarded as an $L^1$-version of Gordon's theorem and it has
a broader range of application. Absence of eigenvalues is then established for quasiperiodic potentials generated by Liouville frequencies and various types of functions such as step functions, H\"older continuous functions and functions with power-type singularities. The proof is based on Gronwall-type a priori estimates for solutions of Schr\"odinger equations.
\end{abstract}

\section{Introduction}
In this paper we study one-dimensional Schr\"odinger operators of the form

\begin{equation}\label{operator}
H = -\frac{d^2}{dx^2} + V ( x ),
\end{equation}
acting on $L^2(\RR)$, with some real-valued $L^1_{\rm loc}$-potential $V$. We will be particularly interested in potentials of the form

\begin{equation}\label{potential}
V ( x ) = V_1 ( x )  + V_2 ( x \alpha + \theta ),
\end{equation}
where we assume that $V_1$ and $V_2$ are $1$-periodic and locally integrable, and $\alpha , \theta \in [0,1)$. If $\alpha = \frac{p}{q}$ is rational, then the potential $V$ is $q$-periodic
and $H$ has purely absolutely continuous spectrum. If $\alpha$ is irrational, then the potential is quasiperiodic and the spectral theory of $H$ is far from trivial; compare
\cite{e,fk1,fk2,fsw,k,ss}.

We want to study the eigenvalue problem for $H$. More precisely, we are interested in methods that allow one to exclude the presence of eigenvalues. A notion that has proved to be useful in
this context is the following. A bounded potential $V$ on $(-\infty, \infty)$ is called a \textit{Gordon potential} if there exist $T_m$-periodic potentials $V^{(m)}$ such that $T_m
\rightarrow \infty$ and for every $m$, $$ \sup_{-2T_m \le x \le 2T_m} |V(x) - V^{(m)}(x)| \le C m^{-T_m} $$ for some suitable constant $C$. It has been shown by Gordon \cite{g} (see also Simon
\cite{s1}) that $H$ has no eigenvalues if $V$ is a Gordon potential. For discrete Schr\"odinger operators, certain variants of this result have been established by Delyon and Petritis
\cite{dp} and by S\"ut\H{o} \cite{s2}; see \cite{d} for a survey of the applications of criteria in this spirit. The applications in the discrete case include in particular results for models that are generated by discontinuous functions, for example, step functions. The interest in such models stems from the theory of one-dimensional quasicrystals; compare \cite{d}. It is clear that in the continuum case, these functions are outside the scope of Gordon's result. This motivates our attempt to find a more general criterion for absence of eigenvalues.

Let us call $V$ a \textit{generalized Gordon potential} if $V \in L^1_{{\rm loc,unif}}(\RR)$, that is,
$$
\|V\|_{1,{\rm unif}} = \sup_{x\in\RR} \int_x^{x+1} |V(x)| dx < \infty
$$
and there exist $T_m$-periodic potentials $V^{(m)}$ such that $T_m \rightarrow \infty$ and for
every $C < \infty$, we have

\begin{equation}\label{ggpcond}
\lim_{m \rightarrow \infty} \exp(C T_m) \cdot \int_{-T_m}^{2T_m} |V(x) - V^{(m)}(x)| dx = 0.
\end{equation}
Clearly, every Gordon potential is a generalized Gordon potential. Our main result is the following:

\begin{theorem}\label{main}
Suppose $V$ is a generalized Gordon potential. Then the operator $H$ in \eqref{operator} has empty point spectrum.
\end{theorem}

As in the classical case \cite{g,s1}, the proof gives the stronger
result that for every energy $E$, the solutions of

\begin{equation}\label{ode}
-u'' (x) + V(x) u(x) = E  u(x)
\end{equation}
do not tend to zero as $|x| \rightarrow \infty$, that is, $|u(x_n)|^2 + |u'(x_n)|^2 \ge D$ for some constant $D > 0$ and a sequence $(x_n)_{n \in \NN}$ which obeys $|x_n| \rightarrow
\infty$ as $n \rightarrow \infty$. Thus there are no $L^2$-solutions since $u\in L^2(\RR)$ would imply $|u(x)|^2+|u'(x)|^2 \rightarrow 0$ as $|x| \rightarrow \infty$ by Harnack's inequality (see \cite{semi}). Note that this uses $V\in L^1_{{\rm loc,unif}}(\RR)$, which also guarantees that the operator $H$ can be defined by form methods or via Sturm-Liouville theory.

Let us now discuss the application of Theorem \ref{main} to quasiperiodic $V$ given by \eqref{potential}. Given some irrational $\alpha \in [0,1)$, we consider its continued fraction
expansion
$$
\alpha = \cfrac{1}{a_1+ \cfrac{1}{a_2+ \cfrac{1}{a_3 + \cdots}}}
$$
with uniquely determined $a_m \in \NN$ and the continued fraction approximants $\alpha_m = p_m/q_m$ defined by

\begin{alignat*}{3}
p_0 &= 0, &\quad    p_1 &= 1,   &\quad  p_m &= a_m p_{m-1} + p_{m-2},\\
q_0 &= 1, &     q_1 &= a_1, &       q_m &= a_m q_{m-1} + q_{m-2};
\end{alignat*}
compare \cite{khin,lang}. Recall that $\alpha$ is called a \textit{Liouville number} if

\begin{equation}\label{liouville}
| \alpha - \alpha_m | \le B m^{-q_m}
\end{equation}
for some suitable $B$, and that the set of Liouville numbers is a dense $G_\delta$-set of zero Lebesgue measure. Given $V$ as in \eqref{potential}, we consider the $q_m$-periodic approximants
$V^{(m)}$ defined by

\begin{equation}\label{approximants}
V^{(m)} ( x ) = V_1 ( x )  + V_2 ( x \alpha_m + \theta ).
\end{equation}

We immediately obtain the following corollary to Theorem \ref{main}.

\begin{coro}
Suppose that for every $C$, we have

\begin{equation}\label{condition}
\lim_{m \rightarrow \infty} \exp(C q_m) \int_{-q_m}^{2q_m} |V_2( x
\alpha + \theta ) - V_2 ( x \alpha_m + \theta)| dx = 0 .
\end{equation}
Then $V$ {\rm (}as given by \eqref{potential}{\rm )} is a generalized Gordon potential and $H$ {\rm (}as given by \eqref{operator}{\rm )} has empty point spectrum.
\end{coro}

Note that for $\alpha,\theta$ fixed, the class of functions $V_2$ obeying \eqref{condition} is a linear space, that is, it is closed under taking finite sums and under multiplication by constants. Moreover, we shall show that condition \eqref{condition} is satisfied, for example, if $V_2$ is a H\"older continuous function, a step function, or a function with power-type singularities, and $\alpha$ is Liouville and $\theta$ arbitrary. 

The organization of this paper is as follows. In Section 2 we establish estimates on solutions of \eqref{ode} which will imply Theorem \ref{main}. The examples for condition (\ref{condition}) are
discussed in Section 3.

\section{Gronwall-Type Solution Estimates and Proof of Theorem \ref{main}}

In this section we study the solutions to the eigenvalue equations associated to two potentials. These two potentials will later be given by a generalized Gordon potential and one of its
approximants. We assume that the solutions have the same initial conditions at $0$. By an a priori estimate for the equivalent first order systems, found by a standard application of Gronwall's lemma (e.g., \cite{Walter}), we can bound the distance of the two solutions by an integral expression involving the distance of the potentials. It is this estimate which allows us to use $L^1$ rather than $L^{\infty}$-bounds in (\ref{ggpcond}). Theorem \ref{main} follows from this bound combined with some useful properties of solutions to periodic eigenvalue equations.

Fix two potentials $W_1 \in L^1_{{\rm loc,unif}}(\RR)$, $W_2 \in L^1_{{\rm loc}}(\RR)$ and some energy $E$ and consider the solutions $u_1,u_2$ of
$$
-u_1''(x) + W_1 (x) u_1 (x) = E u_1 (x), \; -u_2''(x) + W_2 (x) u_2 (x) = E u_2 (x),
$$
subject to
$$
u_1(0) = u_2(0), \; u_1'(0) = u_2'(0), \; |u_1(0)|^2 + |u_1'(0)|^2 = |u_2(0)|^2 + |u_2'(0)|^2 = 1.
$$

\begin{lemma}\label{estimate}
There exists $C = C(\|W_1-E\|_{1,{\rm unif}})$ such that for every $x$,
we have

\begin{equation} \label{u1u2est}
\left \| \left( \begin{array}{c} u_1(x)\\u_1'(x) \end{array} \right) - \left( \begin{array}{c} u_2(x)\\u_2'(x) \end{array} \right) \right\| \le C \exp(C |x| ) \int_{\min(0,x)}^{\max(0,x)}
|W_1(t) - W_2(t)| \cdot |u_2(t)| dt.
\end{equation}
\end{lemma}
\noindent\textit{Proof.} We consider the case $x \ge 0$ (the modifications for $x < 0$ are obvious). We have

\begin{align*}
\left( \begin{array}{c} u_1(x) - u_2(x) \\ u_1'(x) - u_2'(x) \end{array} \right) = & \int_0^x \left( \begin{array}{c} u_1'(t) - u_2'(t) \\ (W_1(t) - E)u_1(t) - (W_2(t) - E)u_2(t) \end{array}
\right) dt\\
= & \int_0^x \left( \begin{array}{c} 0 \\ (W_1(t) - W_2(t)) u_2(t) \end{array} \right) dt \, + \\
& + \int_0^x \left( \begin{array}{c} u_1'(t) - u_2'(t) \\ (W_1(t) - E) (u_1(t) -
u_2(t)) \end{array} \right) dt\\
= & \int_0^x \left( \begin{array}{c} 0 \\ (W_1(t) - W_2(t)) u_2(t) \end{array} \right) dt \, +\\
& + \int_0^x \left( \begin{array}{cc} 0 & 1 \\ W_1(t) - E & 0 \end{array} \right) \cdot \left( \begin{array}{c} u_1(t) - u_2(t) \\ u_1'(t) - u_2'(t) \end{array} \right) dt.
\end{align*}

Hence

\begin{align*}
\left\| \left( \begin{array}{c} u_1(x) - u_2(x) \\ u_1'(x) - u_2'(x) \end{array} \right) \right\| \le & \int_0^x | (W_1(t) - W_2(t))| \cdot |u_2(t)| dt \, +\\
& + \int_0^x \left\| \left( \begin{array}{cc} 0 & 1 \\ W_1(t) - E & 0 \end{array} \right) \right\| \cdot \left\| \left( \begin{array}{c} u_1(t) - u_2(t) \\ u_1'(t) - u_2'(t) \end{array} \right) \right\| dt.
\end{align*}

By Gronwall's lemma \cite{Walter} we therefore get

\begin{align*}
\left \| \left( \begin{array}{c} u_1(x)\\u_1'(x) \end{array} \right) - \left( \begin{array}{c} u_2(x)\\u_2'(x) \end{array} \right) \right\| \le & \int_0^x | (W_1(t) - W_2(t))| \cdot
|u_2(t)| dt \, \times\\
& \times \exp \left( \int_0^x \left\| \left( \begin{array}{cc} 0 & 1 \\ W_1(t) - E & 0 \end{array} \right) \right\| dt \right) .
\end{align*}

Choosing $C$ suitably, the assertion of the lemma follows. \hfill \qedsymbol

\medskip

We see that we can control the difference of the solutions in terms of an integral condition involving the difference of the potentials. The other key ingredient in the proof of Theorem
\ref{main} is the fact that for periodic potentials, we have some knowledge about the norm of the solution vector $(u(x),u'(x))^T$ at certain points $x$. This is made explicit in the following lemma which is essentially well known (particularly in the discrete case \cite{d,dp}).

\begin{lemma}\label{perestimate}
Suppose $W$ is $p$-periodic and $E$ is some arbitrary energy. Then every solution of

\begin{equation}\label{help}
-u''(x) + W(x) u(x) = E u(x),
\end{equation}
normalized in the sense that

\begin{equation}\label{normal}
|u(0)|^2 + |u'(0)|^2 = 1,
\end{equation}
obeys the estimate
$$
\max \left( \; \left\| \left( \begin{array}{c} u(-p)\\u'(-p) \end{array} \right) \right\| , \left\| \left( \begin{array}{c} u(p)\\u'(p) \end{array} \right) \right\| , \left\| \left( \begin{array}{c} u(2p)\\u'(2p) \end{array} \right) \right\| \; \right) \ge \frac{1}{2}.
$$
\end{lemma}
\noindent\textit{Proof.} This follows by the same reasoning as in the discrete case; compare \cite{d,dp}. For the reader's convenience, we sketch the argument briefly. Consider the
solutions $u$ of \eqref{help}. For $x,y \in \RR$, $x < y$, the mapping

\begin{equation}\label{transfer}
M(x,y) : \left( \begin{array}{c} u(x)\\u'(x) \end{array} \right)
\mapsto \left( \begin{array}{c} u(y)\\u'(y) \end{array} \right)
\end{equation}
is clearly linear and depends only on the energy $E$ and the potential on the interval $(x,y)$. Thus, since $W$ is $p$-periodic, we have

\begin{equation}\label{repeat}
M(-p,0) = M(0,p) = M(p,2p) =: M.
\end{equation}

Moreover, by the Cayley-Hamilton theorem, we have

\begin{equation}\label{cht}
M^2 - \tr (M) M + I = 0.
\end{equation}
If $|\tr (M)| \le 1$, we apply this equation to $(u(0),u'(0))^T$ obeying \eqref{normal} and obtain, using \eqref{repeat}, 
$$
\max \left( \;  \left\| \left( \begin{array}{c} u(p)\\u'(p) \end{array} \right) \right\| , \left\| \left( \begin{array}{c} u(2p)\\u'(2p) \end{array} \right) \right\| \; \right) \ge \frac{1}{2},
$$
since $(u(0),u'(0))^T$ has norm one. If $|\tr (M)| > 1$, we apply \eqref{cht} along with \eqref{repeat} to $(u(-p),u'(-p))^T$ and obtain
$$
\max \left( \; \left\| \left( \begin{array}{c} u(-p)\\u'(-p) \end{array} \right) \right\| , \left\| \left( \begin{array}{c} u(p)\\u'(p) \end{array} \right) \right\| \;
\right) \ge \frac{1}{2},
$$
again since the vector $(u(0),u'(0))^T$ has norm one. Put together, we obtain the claimed result. \hfill \qedsymbol

\medskip

We are now in a position to give the

\medskip

\noindent\textit{Proof of Theorem \ref{main}.} Let $V$ be a generalized Gordon potential and let $V^{(m)}$ be the $T_m$-periodic approximants obeying \eqref{ggpcond}. Fix some $m$
and apply Lemma \ref{estimate} with $W_1 = V$ and $W_2 = V^{(m)}$. We obtain

\begin{equation} \label{uumest}
\left \| \left( \begin{array}{c} u(x)\\u'(x) \end{array} \right) - \left( \begin{array}{c} u_m (x)\\u_m'(x) \end{array} \right) \right\| \le C_1 \exp(C_1 |x| ) \int_{\min(0,x)}^{\max(0,x)} |V(t) - V^{(m)}(t)| |u_m(t)| dt,
\end{equation}
where $u$ (resp., $u_m$) solves $-u''(x) + V(x) u(x) = E u(x)$ (resp., $-u_m''(x) + V^{(m)}(x) u_m(x) = E u_m(x)$) and $u,u_m$ are both normalized at the origin and obey the same boundary
condition there. We conclude from \eqref{ggpcond} that $\|V^{(m)}\|_{1,{\rm unif}}$ is bounded in $m$. Thus a second application of Lemma \ref{estimate} with $W_1 = V^{(m)}$ and $W_2 = 0$, noting that the constant in \eqref{u1u2est} only depends on the $L^1_{{\rm loc,unif}}$-norm of $W_1-E$, leads to
$$
\left \| \left( \begin{array}{c} u_m(x)\\u_m'(x) \end{array} \right) - \left(
\begin{array}{c} u_0 (x)\\u_0'(x) \end{array} \right) \right\| \le C_2 \exp(C_2 |x|) \int_{\min(0,x)}^{\max(0,x)} |V^{(m)}(t)| |u_0(t)| dt,
$$
where $C_2$ does not depend on $m$ and $u_0$ is a normalized solution of $-u_0''=Eu_0$. Noting that $u_0$ is exponentially bounded, this gives
$$
|u_m(x)| \le C_3 \exp(C_3 |x|)
$$
with $C_3$ independent of $m$. This and \eqref{uumest} yield
$$
\left \| \left( \begin{array}{c} u(x)\\u'(x) \end{array} \right) - \left( \begin{array}{c} u_m (x)\\u_m'(x) \end{array} \right) \right\| \le C \exp(C|x|) \int_{\min(0,x)}^{\max(0,x)}
|V(t) - V^{(m)}(t)| dt.
$$

By \eqref{ggpcond} we find some $m_0$ such that for $m \ge m_0$, we have
$$
\left \| \left( \begin{array}{c} u(x)\\u'(x) \end{array} \right) - \left(
\begin{array}{c} u_m (x)\\u_m' (x) \end{array} \right) \right\|
\le \frac{1}{4}
$$
for every $x$ with $-T_m \le x \le 2T_m$. Combining this with Lemma \ref{perestimate}, we can conclude the proof. \hfill \qedsymbol

\section{Examples of Generalized Gordon Potentials}

In this section we give examples of functions $V_2$ that obey condition \eqref{condition} for Liouville frequencies $\alpha$ and hence induce quasiperiodic functions $V$ by \eqref{potential} which are generalized Gordon potentials. These will include H\"older continuous functions, step functions, functions with local singularities, and linear combinations thereof.

Let us observe the following:

\begin{prop}
For fixed $\alpha,\theta$, the class of functions $V_2$ obeying \eqref{condition} is a linear space, that is, it is closed under taking finite sums and under multiplication by constants.
\end{prop}
\noindent\textit{Proof.} This is obvious. \hfill \qedsymbol

\medskip

Define for some $1$-periodic function $f$,
$$
\osc_{f,\varepsilon}(x) = \sup_{y,z \in (x - \varepsilon, x + \varepsilon)} | f(y) - f(z) |.
$$
Then we have the following proposition.

\begin{prop}\label{boundex}
{\rm (a)} If there are $0 < \delta , D < \infty$ such that

\begin{equation} \label{osc}
\int_0^1 \osc_{V_2,\varepsilon}(x) dx \le D \varepsilon^\delta
\end{equation}
for all sufficiently small $\varepsilon > 0$, then for every Liouville number $\alpha \in [0,1)$ and every $\theta \in [0,1)$, condition \eqref{condition} is satisfied.\\[1mm]
{\rm (b)} Condition \eqref{osc} holds for all H\"older continuous functions and for all step functions.
\end{prop}
\noindent\textit{Proof.} (a) Fix some $C$. Then by \eqref{liouville} and \eqref{osc}, we have

\begin{align*}
\limsup_{m \rightarrow \infty} \exp(C q_m) \int_{-q_m}^{2 q_m} | & V_2(x \alpha + \theta) - V_2(x \alpha_m + \theta)| dx \le \\
 & \le \limsup_{m \rightarrow \infty} \exp(C q_m) \frac{3 q_m \alpha + 1}{\alpha} \int_0^1 \osc_{V_2,2 q_m |\alpha - \alpha_m|} (x) dx\\
& \le \limsup_{m \rightarrow \infty} \exp(C q_m) \frac{3 q_m \alpha + 1}{\alpha} D (2 q_m |\alpha - \alpha_m|)^\delta\\
& \le \limsup_{m \rightarrow \infty} \exp(C q_m) \frac{3 q_m \alpha + 1}{\alpha} D 2^\delta q_m^\delta B^\delta m^{-\delta q_m}\\
& = 0.
\end{align*}
(b) This is straightforward. \hfill \qedsymbol

\medskip

The class for which \eqref{condition} was established in Proposition \ref{boundex} contains only bounded potentials. We finally provide an example which shows that the use of generalized Gordon potentials allows one to exclude eigenvalues for some unbounded quasiperiodic potentials. We will exhibit some $V_2$ that has an integrable power-like singularity and which satisfies \eqref{condition}, and therefore $H$ defined by \eqref{operator} and \eqref{potential} has empty point spectrum. Note that by linearity this also gives examples with negative singularities and multiple singularities with different values for $\gamma$.

\begin{prop}
Let $0<\gamma<1$ and $V_2(x)$ be the $1$-periodic potential which for $-1/2 \le x \le 1/2$ is given by $V_2(x) = |x|^{-\gamma}$.  Then for every Liouville number $\alpha \in [0,1)$ and every $\theta \in [0,1)$, condition \eqref{condition} is satisfied.
\end{prop}
\noindent\textit{Proof.} For simplicity, we will only establish \eqref{condition} for $\theta = 0$. The calculations for general $\theta$ are similar but slightly more tedious. Start by writing

\begin{equation} \label{split}
\int_{-q_m}^{2q_m} |V_2(\alpha x) - V_2(\alpha_m x)| dx = \frac{q_m}{p_m} \sum_{n=-p_m}^{2p_m} \int_n^{n+1} \left|V_2 \left( \frac{\alpha q_m}{p_m} y \right) - V_2(y) \right| dy
\end{equation}
and

\begin{equation} \label{shift}
\int_n^{n+1} \left| V_2 \left( \frac{\alpha q_m}{p_m} y \right) - V_2(y) \right| dy = \int_0^1 \left| V_2 \left( y+ \left( \frac{\alpha q_m}{p_m} -1 \right) (y+n) \right) - V_2(y) \right|
dy.
\end{equation}

We have $|\frac{\alpha q_m}{p_m}-1| |y+n| \le 2p_m |\frac{\alpha q_m}{p_m} -1| =: \delta < 1/4$ for $m$ sufficiently large and can estimate

\begin{eqnarray} \label{halfest}
& & \int_0^{1/2} \left| V_2 \left(y+ \left( \frac{\alpha q_m}{p_m} -1 \right) (y+n) \right) - V_2(y) \right| dy\\
& & \le C\delta + C \delta^{1-\gamma} + \left| \int_0^{1/2} \left( V_2 \left( y+ \left( \frac{\alpha q_m}{p_m} -1 \right)(y+n) \right) - V_2(y) \right) dy \right|, \nonumber
\end{eqnarray}
where the $\delta^{1-\gamma}$ term arises from the singularity of $V_2$ at $0$, and the monotonicity of $V_2$ in $[0,1/2]$ was used to take the absolute value outside the integral. The integral on the right can be calculated explicitly, which eventually leads to an estimate $C(p_m |\frac{\alpha q_m}{p_m}-1|)^{1-\gamma}$ for its absolute value and thus also for \eqref{halfest}.

In a similar way we get the same estimate for the integral from $1/2$ to $1$ on the right hand side of \eqref{shift}. Inserting into \eqref{split} we finally find
$$
\int_{-q_m}^{2q_m} \left| V_2(\alpha x) - V_2(\alpha_m x) \right| dx \le C p_m^{2-\gamma} \left| \frac{\alpha q_m}{p_m} -1 \right|^{1-\gamma}.
$$
In view of \eqref{liouville} this suffices to imply \eqref{condition}. \hfill \qedsymbol

\end{document}